\documentclass[aps, pra, reprint, showpacs, superscriptaddress]{revtex4-1}
\usepackage{amsmath, amssymb, graphicx, dcolumn, subfigure, CJK, enumerate}
\usepackage{color, hyperref}
\usepackage[all]{hypcap}

\definecolor{MyDarkGreen}{rgb}{0,0.6,0}
\definecolor{MyDarkBlue}{rgb}{0,0,0.8}
\definecolor{MyDarkRed}{rgb}{0.6,0,0.3}
\hypersetup{breaklinks=true,colorlinks=true,plainpages=true,linktocpage=true,linkcolor=MyDarkBlue,citecolor=MyDarkGreen,urlcolor=MyDarkRed,pdfborder={0 0 0},%
pdfauthor={Sang-Kil Son},%
pdftitle={Monte Carlo calculation of ion, electron, and photon spectra of xenon atoms in x-ray free-electron laser pulses}%
}

\newcommand{\figurescale}{1.0}

\begin{document}
\begin{CJK*}{UTF8}{}

\title{Monte Carlo calculation of ion, electron, and photon spectra of xenon atoms in x-ray free-electron laser pulses}

\author{Sang-Kil Son \CJKfamily{mj}(손상길)}
\email{sangkil.son@cfel.de}
\affiliation{Center for Free-Electron Laser Science, DESY, 22607 Hamburg, Germany}

\author{Robin Santra}
\email{robin.santra@cfel.de}
\affiliation{Center for Free-Electron Laser Science, DESY, 22607 Hamburg, Germany}
\affiliation{Department of Physics, University of Hamburg, 20355 Hamburg, Germany}

\date{\today}

\begin{abstract}
When atoms and molecules are irradiated by an x-ray free-electron laser (XFEL), they are highly ionized via a sequence of one-photon ionization and relaxation processes.
To describe the ionization dynamics during XFEL pulses, a rate equation model has been employed.
Even though this model is straightforward for the case of light atoms, it generates a huge number of coupled rate equations for heavy atoms like xenon, which are not trivial to solve directly.
Here, we employ the Monte Carlo method to address this problem and we investigate ionization dynamics of xenon atoms induced by XFEL pulses at a photon energy of 4500~eV.
Charge state distributions, photo-/Auger electron spectra, and fluorescence spectra are presented for x-ray fluences of up to $10^{13}$~photons/$\mu$m$^2$.
With the photon energy of 4500~eV, xenon atoms can be ionized up to +44 through multiphoton absorption characterized by sequential one-photon single-electron interactions.
\end{abstract}

%

\maketitle
\end{CJK*}

\section{Introduction}\label{sec:intro}

The recent advent of x-ray free-electron lasers (XFEL)~\cite{Feldhaus05,Pellegrini04,McNeil10} enables us to explore new frontiers of science~\cite{Piancastelli10}, for example, femtosecond x-ray imaging~\cite{Neutze00,Hajdu00,Gaffney07,Miao08,Mancuso10} and warm dense matter~\cite{Lee03}.
A series of experiments conducted at the Linac Coherent Light Source (LCLS)~\cite{Emma10} have shown how ultraintense and ultrashort x rays interact with various systems: light atom (Ne)~\cite{Young10,Doumy11}, molecule (N$_2$)~\cite{Hoener10,Cryan10,Fang10}, heavy atom (Xe)~\cite{Rudek12}, and solid (Al)~\cite{Vinko12}.

The ionizing XFEL--matter interaction is one of the most fundamental processes that affects all XFEL applications.
As demonstrated theoretically~\cite{Rohringer07} and experimentally~\cite{Young10}, the electronic response to an XFEL pulse is characterized by a sequence of one-photon ionization and relaxation events.
In the x-ray regime, photoabsorption predominantly ionizes an inner-shell electron.
The resulting inner-shell vacancy is filled via radiative (fluorescence) and/or non-radiative (Auger and Coster--Kronig) transitions. 
Then the extremely large number of x-ray photons within an ultrashort XFEL pulse can keep ionizing after or even before these relaxation processes are over~\cite{Moribayashi98,Rohringer07}.
As a result, atoms or molecules become highly ionized after absorbing several photons sequentially.
To describe ionization dynamics, we employ a rate equation model, which demonstrates good agreement with experiments conducted at LCLS~\cite{Young10,Doumy11}.
Tracking populations via rate equations is sufficient to describe the ionization dynamics during XFEL pulses, mainly because the coherence time of current XFEL sources is much shorter than the time scale of population changes.
The ionization dynamics in an XFEL pulse differ from those at a third-generation x-ray synchrotron radiation source, where one-photon absorption is dominant, and from multiphoton strong-field ionization, where many photons are simultaneously absorbed to ionize a single electron.
Understanding radiation damage mechanisms~\cite{Howells09} including ionization dynamics is of central importance for single-shot imaging of individual molecules~\cite{Kai10,Hau-Riege07,Son11a}.

To probe ionization dynamics induced by an XFEL, one can collect all particles generated in the interaction between XFEL and matter. Photoionization and Auger (Coster--Kronig) decay produce electrons, and fluorescence produces photons.
Also highly charged ions are generated via multiphoton multiple ionization.
It is possible to simultaneously measure all those particles by means of the CFEL--ASG Multi-Purpose (CAMP) instrument~\cite{Struder10}, which has been successfully applied to a study of the XFEL--heavy atom interaction~\cite{Rudek12} and to single-shot imaging experiments~\cite{Chapman11,Seibert11,Barty12,Koopmann12,Johansson12}.

Ionization of heavy atoms irradiated by XFEL pulses has attracted considerable attention, not only because the heavy atom has many electrons to be ionized but also because it has a rich manifold of ionization channels involving complex inner-shell decay cascades.
A recent study proposes a resonance-enabled x-ray multiple ionization mechanism for heavy atoms to reach high charge states beyond those expected from the straightforward sequential ionization model~\cite{Rudek12}.
Also, ionization dynamics of heavy atoms embedded in macromolecules deliver a novel way to determine macromolecular structure from femtosecond nanocrystallography data using XFEL~\cite{Son11e}.
The theoretical treatment of the XFEL--heavy atom interaction is challenging because its dynamics involve a huge number of possible pathways.
For example, the simple rate equation model for Xe $M$-shell ionization requires more than one million coupled rate equations~\cite{Rudek12}.
This number becomes even much larger when resonantly excited Rydberg states are taken into account.
Therefore, it is necessary to develop a computational tool that can handle heavy atoms and provide full information on ions, electrons, and photons.

In the present work, we employ the Monte Carlo method to solve the large number of coupled rate equations.
There have been extensive studies using the Monte Carlo procedure for decay pathways of an inner-shell vacancy produced by x-ray synchrotron or electron capture~\cite{Carlson65a,Krause67,Charlton81,Mukoyama85,Mukoyama86,Mukoyama87,Pomplun87,Mirakhmedov88,Opendak90,Mohammedein93,El-Shemi97,Abdullah03,El-Shemi05}.
In particular, the inner-shell decay process for heavy atoms such as iodine and xenon~\cite{Charlton81,Pomplun87,Mukoyama86,Mukoyama87} has brought much interest because of their relevance for medical applications~\cite{Pomplun87,Kassis11}.
In conventional implementations of the Monte Carlo method, atomic data for multiple-hole ions are usually assumed to be the same as for the singly-ionized atom~\cite{Mukoyama86} or they are scaled from the singly-ionized atom according to the number of valence electrons~\cite{Larkins71}.
To the best of our knowledge, no Monte Carlo calculation has been done using atomic data that are individually calculated for all  possible multiple-hole configurations.
In the current implementation, we perform electronic structure calculations individually for all possible configurations in order to obtain a whole set of atomic data.
This is important for ionization dynamics in XFEL radiation because of the production of a broad range of charge states.
The higher the ionic charge, the greater are the deviations from the singly-ionized atom.
As the charge state increases, some Auger transitions become energetically forbidden~\cite{Mirakhmedov88}.
Thus, the detailed electronic structure for each configuration matters to atomic data calculations and eventually to ionization dynamics simulations.
Another distinct aspect of the Monte Carlo implementation described in this paper is the availability of temporal information on electronic dynamics.
In ordinary Monte Carlo simulations of electronic decay cascades, the time variable has not been of main concern, and in many cases decay rates are simply given by relative rates.
However, it is important to know details of ionization dynamics, especially when their time scale is comparable with the time scale of nuclear dynamics~\cite{Dunford12a}.

The structure of this paper is as follows.
In Sec.~\ref{sec:theory}, we describe the rate equation model for ionization dynamics and present a Monte Carlo implementation for solving a large set of coupled rate equations.
In Sec.~\ref{sec:result}, we present atomic data of Xe and time-dependent ionization pathways from Monte Carlo simulations.
We discuss ionization dynamics of Xe in intense x-ray pulses at a photon energy of 4500~eV by analyzing ion, electron, and photon spectra.
We conclude with a summary and outlook in Sec.~\ref{sec:conclusion}.

\section{Theory and numerical details}\label{sec:theory}

\subsection{Ionization dynamics}
To simulate ionization dynamics in intense x-ray pulses, we employ a rate equation approach based on sequential one-photon ionization and relaxation steps.
This rate equation model was introduced in connection with XFEL--atom interactions by Rohringer and Santra~\cite{Rohringer07}, extended to x-ray scattering dynamics~\cite{Son11a}, generalized to arbitrary elements~\cite{xatom}, and has been successfully applied to explain recent LCLS experiments~\cite{Young10,Doumy11,Rudek12}.
A rate equation model for non-local thermal equilibrium plasma~\cite{Chung05,Chung07} has been applied to a warm dense matter study at LCLS~\cite{Vinko12}.

Here, we summarize procedures underlying the rate equation model.  
For interested readers, theoretical background~\cite{Santra09} and detailed descriptions~\cite{Son11a} are available.
For a given atom, we construct all possible electronic configurations $\lbrace I \rbrace$ that may be formed by removing zero, one, or more electrons, from the neutral ground configuration.
The orbital structures are optimized with the Hartree--Fock--Slater (HFS) method for each configuration.
We include all possible one-photon ionization and relaxation processes for each configuration, i.e., the subshell photoionization cross sections for a given photon energy, Auger (Coster--Kronig) decay rates, and fluorescence rates are calculated for every single configuration.
We calculate shake-off branching ratios, based on the sudden change approximation~\cite{Krause64}.
In the present work, shake-off processes are included for all photo-induced processes of neutral Xe.
The calculated cross sections and rates serve as input parameters for a set of rate equations of the form,
\begin{equation}\label{eq:rate_equation}
\frac{d}{dt} P_I(t) = \sum_{I' \neq I}^\text{all config.} \left[ \Gamma_{I' \rightarrow I} P_{I'}(t) - \Gamma_{I \rightarrow I'} P_I(t) \right],
\end{equation}
where $P_I$ is the population of the $I$th configuration, and $\Gamma_{I \rightarrow I'}$ is the rate for a transition from the configuration $I$ to the configuration $I'$.

For the heavy atom case, the numbers of configurations and processes involved in ionization dynamics are very large.
For example, an x-ray photon of 4500~eV can ionize the $M$-, $N$-, and $O$-shells of Xe.
The number of all possible electronic configurations constructed from these ionizations is 1,120,581, which is equal to the number of coupled rate equations to be solved, and the number of all possible processes under consideration is 43,221,650.
Therefore we need to propagate in time a matrix of approximately one million by one million with $\sim$40 million nonzero elements, in order to simulate, within the rate equation model, ionization dynamics of Xe exposed to 4500-eV XFEL pulses.
To avoid a direct time-propagation solution of this huge matrix, we develop a Monte Carlo approach that efficiently solves the rate equations, as an extension of the \textsc{xatom} toolkit~\cite{xatom}.

\subsection{Monte Carlo implementation}\label{sec:MC}

Our Monte Carlo description of ionization dynamics may be summarized as follows.
For one realization (a Monte Carlo trajectory), a given atom undergoes a sequence of photoionization and relaxation events during the time propagation and eventually ends up in a final charge state.
There are many pathways to reach the same final charge state.
Each pathway consists of many steps of photoionization and relaxation, and those steps are stochastically determined.
Probabilities are calculated at a given time and the time increment is optimally determined during the time propagation. 
If it is no longer possible to proceed to a further process, the time propagation ends and this trajectory is complete.
We repeat this procedure for many trajectories to form a sufficiently large statistical ensemble.

In the direct time-propagation solution, configuration populations are given by a fractional number representing the probabilities for all individual configurations.
On the other hand, for each realization of the Monte Carlo implementation, configuration populations are given by either zero or one, thus following a specific pathway of configuration changes induced by photoionization and relaxation events.
After running many trajectories, we obtain ensemble-averaged configuration populations, which are ideally the same as the configuration populations obtained by the direct solution.

Here is a more detailed description of our Monte Carlo implementation.
\begin{enumerate}[(a)]
\item Choose an initial value of $\Delta t$.
This is also used for the maximum value of $\Delta t$.

\item Set up an initial configuration $I$, which is usually given by the ground configuration of a neutral atom.
Set an initial value for the time $t$.

\item Calculate transition probabilities, $\lbrace p_k \rbrace$ for $1 \leq k \leq N_\text{proc}$.
Here, $k$ indicates an index for the transition process of $I \rightarrow I'$, and $N_\text{proc}$ is the number of all possible processes from $I$.
\[
p_k = 
\left\{
\begin{array}{l}
\sigma^\text{P}_k J(t) \Delta t \quad \text{for photoionization}\vspace{.1in},
\\
\Gamma^\text{A}_k \Delta t \quad \text{for Auger (Coster--Kronig) decay}\vspace{.1in},
\\
\Gamma^\text{F}_k \Delta t \quad \text{for fluorescence},
\end{array}
\right.
\]
where $J(t)$ is the photon flux of the x-ray pulse at a given time $t$, $\sigma^\text{P}$ is the photoionization cross section, $\Gamma^\text{A}$ is the Auger (Coster--Kronig) rate, and $\Gamma^\text{F}$ is the fluorescence rate.

\item Construct a table of processes, $\lbrace T_k \rbrace$.
$T_0 = 0$ and $T_k = \sum_{k'=1}^{k} p_{k'}$ for $1 \leq k \leq N_\text{proc}$.
Here, $T_{N_\text{proc}}$ gives the total probability to proceed to one of the $N_\text{proc}$ processes, whereas $1 - T_{N_\text{proc}}$ gives the probability to remain in configuration $I$.

\item Choose a random number, $r \in [ 0, 1 ]$.  
If $T_{k-1} < r \leq T_{k}$, then it proceeds to the $k$th process and the new configuration becomes $I'$.
For Auger (Coster--Kronig) decay and photoionization processes, the electron count corresponding to its kinetic energy bin is increased by one.
For fluorescence, the photon count corresponding to its emitted photon energy bin is increased by one.

\item Adjust $\Delta t$ according to the calculated $T_{N_\text{proc}}$.
It must satisfy $T_{N_\text{proc}} \ll 1$ and must not be larger than the initial $\Delta t$.
Then increase the time variable $t$ by $\Delta t$.

\item Repeat (c)--(f) as long as $T_{N_\text{proc}} \neq 0$.
If $T_{N_\text{proc}} = 0$, there will be no further process.
This finishes one Monte Carlo trajectory of the time-propagation calculation.
The final charge state count is increased by one.

\item Run many trajectories until the results are converged.

\item The counts of charge state, electron energy, and photon energy are divided by the number of trajectories.
These histograms correspond to the ion, electron, and photon spectra, respectively.
\end{enumerate}

We use the following computational parameters.
For (a), the initial and maximum value of $\Delta t$ is 10 attoseconds.
For (f), $\Delta t$ is chosen such that $T_{N_\text{proc}} = 0.1$ during the time propagation. 
For the convergence criterion of (h), we check charge-state populations for every 100 trajectories.
In practice, 10,000 to 30,000 trajectories are carried out to obtain convergency of $10^{-4}$ for all charge-state populations.
These Monte Carlo results fully agree with the direct solution to within an accuracy of $10^{-3}$ discrepancies.
There is a tremendous reduction in the computational time.
The direct solution takes about 1,760 minutes for a Xe-atom ionization dynamics calculation with 16,000 time steps on the lab workstation, whereas the Monte Carlo implementation takes only 7 minutes with about 20,000 trajectories to get converged results.

\section{Results and discussion}\label{sec:result}

\subsection{Atomic data}\label{sec:atomic_data}
\begin{figure}
\includegraphics[scale=\figurescale]{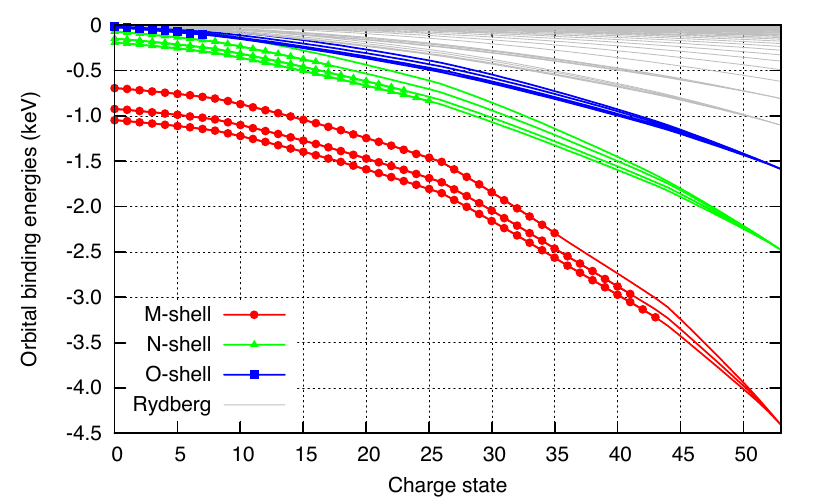}
\caption{\label{fig:orbital}%
(Color online) Orbital binding energies of the ground configuration of Xe and its charge states.
The symbols (circle, triangle, and square) indicate that the corresponding subshell contains at least one electron.}
\end{figure}

Figure~\ref{fig:orbital} shows orbital binding energies of the ground configuration of Xe$^{q+}$ as a function of the charge state $+q$.
The red lines ($M$-shell) from the bottom correspond to the $3s$, $3p$, and $3d$ subshells.
The green lines ($N$-shell) correspond to the $4s$, $4p$, and $4d$ subshells, and the blue lines ($O$-shell) correspond to the $5s$ and $5p$ subshells.
The dots with circles, triangles, and rectangles, indicate that the corresponding subshell is fully or partially filled with electrons for given charge states.
Here are some examples of the ground configuration of Xe$^{q+}$,
\begin{align*}
\text{Xe}^{0+}:\ & 1s^{2}2s^{2}2p^{6}3s^{2}3p^{6}3d^{10}4s^{2}4p^{6}4d^{10}5s^{2}5p^{6},
\\
\text{Xe}^{8+}:\ & 1s^{2}2s^{2}2p^{6}3s^{2}3p^{6}3d^{10}4s^{2}4p^{6}4d^{10},
\\
\text{Xe}^{26+}:\ & 1s^{2}2s^{2}2p^{6}3s^{2}3p^{6}3d^{10},
\\
\text{Xe}^{44+}:\ & 1s^{2}2s^{2}2p^{6}.
\end{align*}
As shown in Fig.~\ref{fig:orbital}, the photon energy of 4500~eV is well above all ionization potentials of $M$-, $N$-, and $O$-shell electrons for all charge states of Xe.
No resonance transition is expected with this photon energy.
Thus one can expect Xe$^{44+}$ as the maximum charge state after multiphoton multiple ionization by x rays of 4500~eV, if the x-ray photon fluence is high enough to remove all $n$$\geq$3 electrons via a sequence of photoionization and relaxation processes.

Table~\ref{table:rate} compares fluorescence, Auger, and Coster--Kronig rates with semi-empirical calculations~\cite{McGuire72a} for $M$-shell single-hole configurations of Xe.
Semi-empirical calculations employ transition energies from experiments, whereas the present method computes them from HFS orbital energies.
For this reason, most semi-empirical calculations consider only single-hole or double-hole configurations.
However, we emphasize that for the present Monte Carlo simulations all multiple-hole configurations are individually calculated with the HFS method.
Even though the rates presented in this table are summed over subshells $X$ and $Y$, all transitions to individual subshells are calculated for the present Monte Carlo simulations and those numbers are comparable with the extensive table~\cite{McGuire72} for the single-hole configurations.
The semi-empirical calculations include relativistic changes in the transition energies via $jj$-coupling~\cite{Ibyari72}.
On the other hand, the present calculations based on the nonrelativistic HFS method do not include fine-structure splittings (for example, between $M_2$ and $M_3$, or between $M_4$ and $M_5$).
Without relativity, $M_2$--$M_3 X$ Coster--Kronig transitions are energetically impossible~\cite{Chen83a,Crasemann84}, so this transition is completely absent in the present method as shown in Table~\ref{table:rate}.
In spite of these limitations of the HFS method, this comparison shows good agreement.

\begin{table}
\caption{\label{table:rate}%
Comparison of fluorescence ($M_i$--$X$), Auger ($M_i$--$XY$), and Coster--Kronig ($M_i$--$M_j X$) rates for $M$-shell single-hole configurations of Xe.
The rates are given in atomic units.
SE represents semi-empirical calculations~\cite{McGuire72a}.
$M_{23}$ and $M_{45}$ in the initial state are averaged over different $j$ and those in the final state are summed.
}
\begin{ruledtabular}
\begin{tabular}{lrr}
Transition & SE~\cite{McGuire72a} & Present \\
\hline
$3s$ hole \\
$M_1$--$X$			& $1.76\times10^{-4}$ & $1.73\times10^{-4}$ \\
$M_1$--$XY$			& $2.06\times10^{-2}$ & $1.85\times10^{-2}$ \\
$M_1$--$M_{23}X$		& $2.78\times10^{-1}$ & $4.76\times10^{-1}$ \\
$M_1$--$M_{45}X$		& $7.59\times10^{-2}$ & $8.98\times10^{-2}$ \\
\hline
$3p$ hole \\
$M_{23}$--$X$		& $1.45\times10^{-4}$ & $1.62\times10^{-4}$ \\
$M_{23}$--$XY$		& $2.18\times10^{-2}$ & $2.10\times10^{-2}$ \\
$M_{2}$--$M_{3}X$	& $1.83\times10^{-3}$ & -- \phantom{$10^{-3}$} \\
$M_{23}$--$M_{45}X$	& $1.70\times10^{-1}$ & $2.06\times10^{-1}$ \\
\hline
$3d$ hole \\
$M_{45}$--$X$		& $6.75\times10^{-5}$ & $1.03\times10^{-5}$ \\
$M_{45}$--$XY$		& $2.49\times10^{-2}$ & $2.26\times10^{-2}$ \\
\end{tabular}
\end{ruledtabular}
\end{table}

The photoionization cross sections for neutral Xe are compared between the relativistic method and the present method in Table~\ref{table:pcs}.
The relativistic results are based on the Dirac--Fock--Slater method~\cite{Band79} and summed over different total angular momentum $j$ to make a comparison with the nonrelativistic case.
For this case, the present results are in excellent agreement with the relativistic results.

\begin{table}
\caption{\label{table:pcs}%
Comparison of photoionization cross sections (in kilobarns) for neutral Xe.
The Dirac--Fock--Slater (DFS) results~\cite{Band79} are calculated at 4509~eV and the present results are at 4500 eV.
The DFS results are summed over different total angular momentum $j$.
}
\begin{ruledtabular}
\begin{tabular}{lrr}
Subshell & DFS~\cite{Band79} & Present \\
\hline
3$s$ &  7.99 &  7.99 \\
3$p$ & 25.33 & 24.20 \\
3$d$ & 16.37 & 16.22 \\
4$s$ &  1.84 &  1.82 \\
4$p$ &  4.96 &  4.73 \\
4$d$ &  2.58 &  2.57 \\
5$s$ &  0.28 &  0.27 \\
5$p$ &  0.54 &  0.52 \\
\end{tabular}
\end{ruledtabular}
\end{table}

Finally, Table~\ref{table:width} lists decay widths of single-hole configurations of Xe.
The widths are calculated by the sum of fluorescence, Auger, and Coster--Kronig rates.
The present results are compared with semi-empirical calculations~\cite{McGuire71,McGuire72a,McGuire74} and recommended values from various experiments and theories~\cite{Campbell01}.
Bearing in mind some limitations of the nonrelativistic treatment for heavy atoms, the present results are in fair agreement with other available values.

\begin{table}
\caption{\label{table:width}%
Comparison of the decay widths (in eV) of single-hole configurations of Xe.
EXP represents recommended values from various experiments and theories~\cite{Campbell01}.
SE refers to semi-empirical calculations~\cite{McGuire71,McGuire72a,McGuire74}.
EXP and SE values are averaged over different $j$.
}
\begin{ruledtabular}
\begin{tabular}{lrrr}
Hole & EXP\footnote{Ref.~\cite{Campbell01}} & SE\footnote{Refs.~\cite{McGuire71,McGuire72a,McGuire74}} & Present \\
\hline
4$d^{-1}$ &  0.09 &  0.08 &  0.05 \\
4$p^{-1}$ &    -- &  2.56 &  2.42 \\
4$s^{-1}$ &  2.60 &  5.49 &  6.93 \\
3$d^{-1}$ &  0.60 &  0.68 &  0.62 \\
3$p^{-1}$ &  4.30 &  5.26 &  6.15 \\
3$s^{-1}$ & 10.60 & 10.18 & 16.05 \\
2$p^{-1}$ &  2.90 &  2.95 &  2.84 \\
2$s^{-1}$ &  2.00 &  4.08 &  4.06 \\
1$s^{-1}$ & 11.50 &       & 11.75 \\
\end{tabular}
\end{ruledtabular}
\end{table}

\subsection{Ionization pathways}
\begin{figure}
\includegraphics[scale=\figurescale]{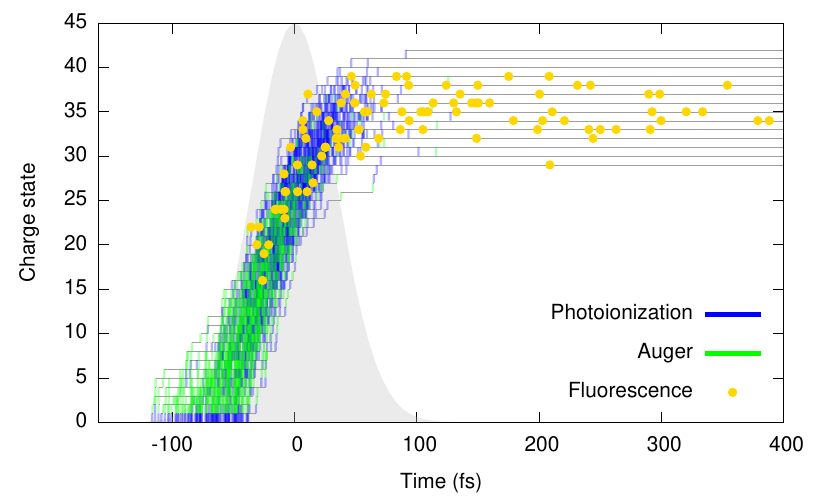}
\caption{\label{fig:trajectory}%
(Color online) Pathways of 100 exemplary trajectories of ionization dynamics of Xe at 4500~eV, 80~fs FWHM, and 5$\times$10$^{12}$~photons/$\mu$m$^2$.
Multiphoton multiple ionization is described by a sequence of one-photon ionization (blue), Auger decay (green), and fluorescence (yellow).
The gray background shows the Gaussian pulse profile of 80~fs FWHM.}
\end{figure}

By analyzing Monte Carlo trajectories, one can retrieve useful information on ionization pathways during XFEL pulses.
Figure~\ref{fig:trajectory} shows 100 exemplary trajectories that are randomly chosen out of 22,200 trajectories of Xe at 4500~eV.
The pulse duration is 80~fs full-width-at-half-maximum (FWHM) and the fluence is 5$\times$10$^{12}$~photons/$\mu$m$^2$.
The blue and green bars represent photoionization and Auger (Coster--Kronig) decay, respectively, and the yellow dots indicate fluorescence.
The bar colors are transparent, so darker colors mean that it is more probable to pass through those pathways.
The ionization dynamics are obviously initiated by $M$-shell one-photon ionization, as shown by the blue bars between charge states zero and one.
This is followed by a series of Auger decays, as shown by the green area above the initial blue bars, i.e., an Auger cascade after one-photon absorption~\cite{Carlson66}.
Decay pathways after one-photon 3$d$-shell ionization of Xe have been studied experimentally~\cite{Tamenori02,Suzuki11} and theoretically~\cite{Jonauskas03,Kochur04,Kochur08}.
Note that the time scale of the Auger cascade ranges from 10 to 100~fs, as depicted in Fig.~\ref{fig:trajectory}.
Thus, more than one photon can be absorbed before the Auger cascade ends, as shown by sparse blue bars inside the green area, thus opening up new channels for ionization.
As the charge state goes up, Auger decays become less likely~\cite{Young10}, so photoionization becomes dominant for further ionization around the peak of the pulse profile.
Fluorescence typically occurs at high charge states, when its rate overcomes the Auger rate. 
The final charge states are formed in the middle of the latter half of the pulse.
At the center of the pulse the charge states around +20 to +30 are formed, and at the end of the pulse they are distributed around +30 to +40.
The pulse-weighted time-averaged charge state is +24 for this fluence case.

\subsection{Ion, electron, and photon spectra}
\begin{figure}
\includegraphics[scale=\figurescale]{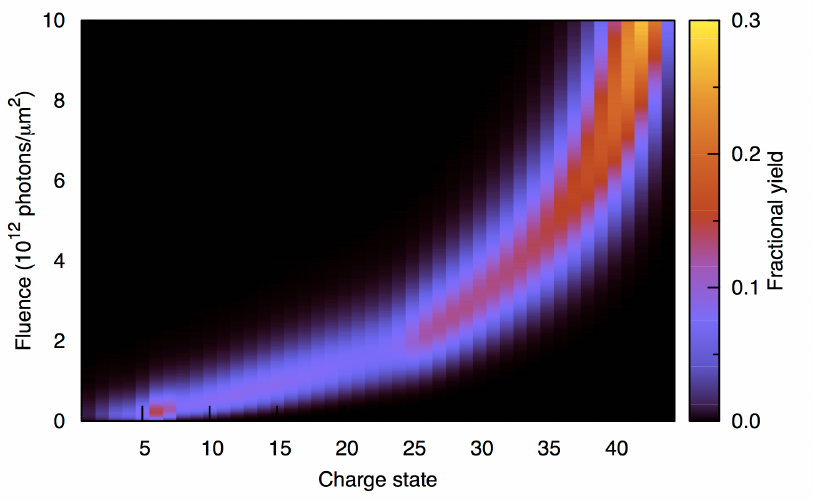}
\caption{\label{fig:ion}%
(Color online) Charge state distribution of Xe at 4500~eV as a function of the fluence.}
\end{figure}

\begin{figure}
\includegraphics[scale=\figurescale]{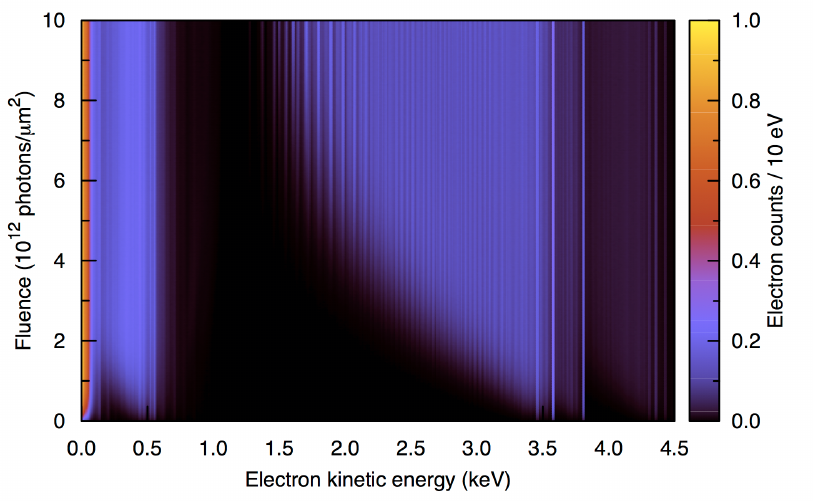}
\caption{\label{fig:electron}%
(Color online) Electron spectra of Xe at 4500~eV as a function of the fluence.
The photoelectrons are above 1250~eV and the Auger (Coster--Kronig) electrons are below 1250~eV.}
\end{figure}

\begin{figure}
\includegraphics[scale=\figurescale]{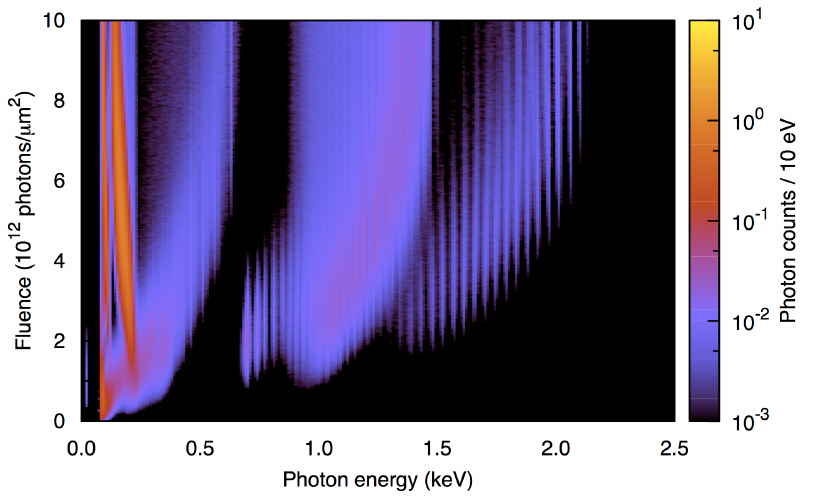}
\caption{\label{fig:photon}%
(Color online) Fluorescence spectra of Xe at 4500~eV as a function of the fluence.
The peak assignments are explained in the text.}
\end{figure}

After calculating all 40 million atomic data parameters for one million configurations, rate equations for given XFEL parameters are solved in the Monte Carlo fashion described in Sec.~\ref{sec:MC}.
From the Monte Carlo simulations, we investigate ion, electron, and photon spectra by counting the final charge states, and photo-/Auger electrons and emitted photons in the energy bins.
The pulse envelope is Gaussian and the pulse duration is 80~fs FWHM.
In the regime of sequential ionization dynamics, the spectra are largely insensitive to the temporal pulse shape and the pulse duration~\cite{Rohringer07}.
The maximum fluence used is $10^{13}$~photons/$\mu$m$^2$.
Even at the peak intensity of this fluence, the inverses of all photoionization rates are longer than 2.6 femtoseconds.
The typical bandwidth of current XFEL sources operating in the hard x-ray regime is about 1\% of the photon energy at LCLS~\cite{Young10,Doumy11,Rudek12} or tens of eV at SACLA in Japan~\cite{Ueda}.
If the XFEL bandwidth is given by 45~eV FWHM for a photon energy of 4500~eV, then the coherence time is about 40 attoseconds.
Thus the coherence time is much shorter than the time scale of the fastest photoionization process, which warrants the use of the rate equation model in this regime.

Figure~\ref{fig:ion} shows the charge state distribution of Xe at 4500~eV after the x-ray pulse is over and all decay processes are completed.
The vertical axis is the fluence varying from zero to $10^{13}$~photons/$\mu$m$^2$, and the color indicates the fractional yield of a given charge state.
The fractional yields are normalized such that the sum of all populations of ions and neutral atom is one.
Therefore, one can observe that the total yield of ions is increasing as the fluence increases.
For clarity, the population of neutral Xe is not shown in the plot.
Near zero fluence, which corresponds to the synchrotron radiation limit, the charge state distribution peaks around +6 and +7.
This distribution is due to the decay cascades of $M$-shell single vacancies~\cite{Mukoyama87,Saito92a,Kochur94}.
When the fluence increases, the charge state distribution is shifted to higher charge states.
The maximum charge state is +44, where all $M$-, $N$-, and $O$-shell electrons are ionized.
When the photon energy is not enough to ionize electrons by absorbing one photon, two-photon ionization may occur~\cite{Doumy11,Sytcheva12a} or resonantly excited states can play a role in generating higher charge states~\cite{Rudek12}.
As shown in Fig.~\ref{fig:orbital}, the 4500-eV photon energy is large enough to ionize all electrons above the $L$-shell via one-photon absorption for all charge states. 
Therefore contributions from direct two-photon ionization and resonant pathways are negligible at 4500~eV.
The maximum charge state of +44 can be reached via a sequence of one-photon processes.

In Fig.~\ref{fig:electron}, we plot electron spectra of Xe at 4500~eV, including both photoelectrons and Auger (Coster--Kronig) electrons.
The vertical axis is the fluence and the horizontal axis is the electron kinetic energy spaced by the 10-eV width of the energy bins.
Below $\sim$0.5~keV, Auger (Coster--Kronig) electrons appear, and photoelectrons have higher kinetic energy in the range from 1.5 to 4.0~keV.
There are three peak lines from photoelectrons: 3.5, 3.6, and 3.8~keV, corresponding to photoionization from $3s$, $3p$, and $3d$ of neutral Xe, respectively.
From x-ray atomic data for neutral Xe~\cite{Thompson01}, those lines are located at 3.4, 3.5, and 3.8~keV.
Shake-off satellite structures in the electron spectra are not included in our calculations.
The fine satellite structures shown in Fig.~\ref{fig:electron}, especially in the photoelectron spectra, are due to different charge states and individual configurations.
With increasing the charge state, the ionization threshold becomes larger and accordingly the photoelectron kinetic energy becomes smaller.
Therefore, as the fluence increases, higher charge states are formed and the photoelectron spectra extend to lower energies.

\begin{figure}
\includegraphics[scale=\figurescale]{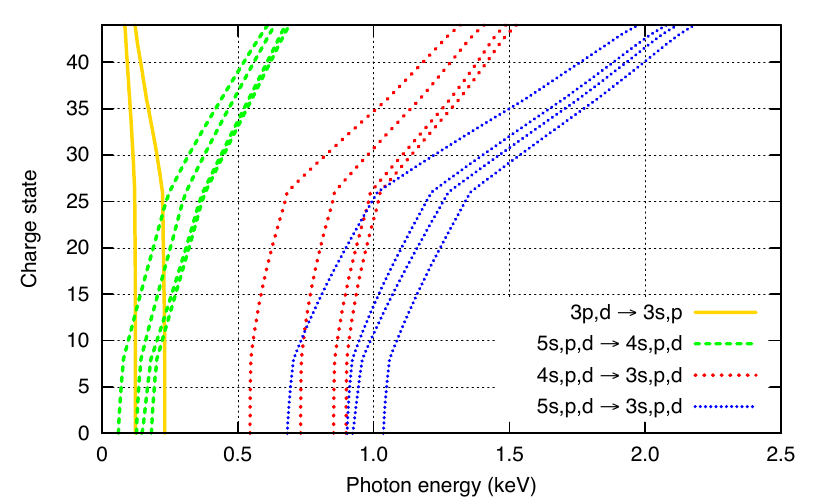}
\caption{\label{fig:trans_E}%
(Color online) Selected fluorescence energies of Xe as a function of the charge state.}
\end{figure}

Figure~\ref{fig:photon} shows fluorescence spectra of Xe at 4500~eV.
Note that the width of the energy bins is 10~eV and the color bar is in the logarithmic scale.
The comb structure comes from different charge states and the fringes are due to different electronic configurations.
The photon spectra can be grouped according to different transition channels: i) strong lines below $\sim$300~eV, ii) a plume from 0~keV to 0.5~keV, iii) a comb structure plus a cloud of lines from 0.7~keV to 1.5~keV, and iv) comb lines from 1.3~keV to 2.2~keV.
To assign those parts, we plot in Fig.~\ref{fig:trans_E} fluorescence energies for several transition channels as a function of the charge state.
These energies are calculated from orbital binding energies of the ground configurations for given charge states.
When the charge state is increased, transition energies between different shells (different quantum number $n$) are increased because the energy levels for $n$ are approximately proportional to the square of the charge state.
On the other hand, transition energies between subshells of the same $n$ (but different quantum number $l$) are decreased because electronic screening becomes less for higher charge states.
Therefore, the four different groups in Fig.~\ref{fig:photon} can be assigned as follows:
i) $n$=3 to $n$=3,
ii) $n$=5 to $n$=4,
iii) $n$=4 to $n$=3, and
iv) $n$=5 to $n$=3.
Contrary to the electron spectra, the photon spectra extend to increasing energies as the fluence increases.
If the energy resolution of the photon spectra is better than 10~eV as used in Fig.~\ref{fig:photon}, then it is possible to observe these comb structures in the fluorescence lines and to assign them to individual charge states, which was recently demonstrated at LCLS in the case of solid aluminum~\cite{Vinko12}.

\section{Conclusion}\label{sec:conclusion}
In this paper, we have implemented a Monte Carlo strategy for solving a rate equation model describing ionization dynamics induced by intense x-ray pulses.
Photoionization cross sections, Auger (Coster--Kronig) rates, and fluorescence rates are calculated for all possible multiple-hole configurations.
Based on the pre-calculated table of all atomic data, Monte Carlo sampling finds probable pathways to reach the final charge states.
Using the \textsc{xatom} toolkit extended by this Monte Carlo method, we have investigated ionization dynamics of Xe in 4500-eV XFEL pulses.
Detailed ionization and relaxation pathways have been depicted as a function of time.
We have plotted the charge state distribution, photo-/Auger electron spectra, and fluorescence spectra as a function of fluence, whose range is experimentally accessible.
Near the upper end of this range, Xe at 4500~eV can be ionized up to +44 via a sequence of one-photon ionization and relaxation processes.

Finally, we would like to briefly describe perspectives for further development.
First, the current Monte Carlo implementation could be called ``brute-force'' since it calculates atomic data for all possible configurations and physical processes.
The Monte Carlo sampling is applied for solving rate equations, but not for calculating atomic data.
It is plausible to integrate both atomic data calculation and search for probable pathways into the Monte Carlo procedure.
In this way, atomic data are computed only when they are required.
Second, it is important to include bound-to-bound photoexcitation processes in the model, which may play a crucial role in ionization dynamics at certain conditions.
Because ionization thresholds have a broad range according to charge states and because XFEL pulses typically have a broad bandwidth, resonance conditions may be easily satisfied.
However, treatment of singly or multiply excited states is theoretically challenging, and moreover, inclusion of all these additional excited-state configurations is numerically demanding.
Third, relativistic effects are not considered in the present work, because $M$-, $N$-, and $O$-shell ionization of Xe may be described reasonably well within the HFS model.
It is conceivable that spin-orbit energy splittings open additional decay channels, but the configurational space will expand substantially when all splittings are taken into account.
Work towards overcoming these challenges is in progress.

\begin{acknowledgments}
We thank Stefan Pabst, Mohamed El Amine Madjet, Benedikt Rudek, Daniel Rolles, and Artem Rudenko for helpful discussions.
\end{acknowledgments}


\end{document}